\documentstyle[epsf,12pt]{article}

\begin{document}
\renewcommand{\baselinestretch}{1.5}

\newcommand\beq{\begin{equation}}
\newcommand\eeq{\end{equation}}
\newcommand\bea{\begin{eqnarray}}
\newcommand\eea{\end{eqnarray}}

\newcommand\rRo{\rho_{R0}^A}
\newcommand\rLo{\rho_{L0}^A}
\newcommand\rR{\rho_{R}^A}
\newcommand\rL{\rho_{L}^A}

\newcommand\al{\alpha}
\newcommand {\dlt}{ \frac{\delta}{\pi}}
\newcommand {\dlts}{\frac{\delta^2}{\pi^2}}

\newcommand\pisig{\Pi_{\sigma i }}
\newcommand\pisigm{\Pi_{I\sigma i }}
\newcommand\sumsig{\sum_{\sigma i }}
\newcommand\psig{p_{\sigma i }}
\newcommand\xsig{x_{\sigma i }}
\newcommand\xmsig{x_{-\sigma j }}
\newcommand\sumI{\sum_{I}^{N}}
\newcommand\sqpi{{\sqrt\pi}}
\newcommand\sumi{\sum_{i=1}^{N}}
\newcommand\sumj{\sum_j}
\newcommand\sumjN{\sum _{j=1} ^N}
\newcommand\sumJ{\sum_J}

\newcommand\expo{e^{iS(\{x_{+i}\}, \{x_{-i}\})}}
\newcommand\expon{e^{-iS(\{x_{+i}\}, \{x_{-i}\})}}

\newcommand\tp{\tilde\phi}
\newcommand\dpi{\delta/\pi}
\newcommand\dtpi{\delta/2\pi}
\newcommand\ddpi{\delta^2/\pi^2}
\newcommand\px{\partial_x}
\newcommand\pt{\partial_t}
\newcommand\prl{Phys. Rev. Lett.}
\newcommand\prb{Phys. Rev. {\bf B}}
\hfill MRI-PHY/P990308

\centerline{\bf Results from Bosonisation for Resonant Tunneling through} 
\centerline{\bf a Quantum Dot in an Aharanov-Bohm Ring}
\vskip 1 true cm

\centerline{P. Durganandini \footnote{{\it e-mail address}: 
pdn@physics.unipune.
ernet.in}}
\centerline{\it Department  of Physics, Pune University, }
\centerline{\it Pune 411 007, India.}
\vskip .5 true cm

\centerline{Sumathi Rao \footnote{{\it e-mail
address}: sumathi@mri.ernet.in}}  
\centerline{\it Mehta Research Institute, Chhatnag Road, Jhunsi,}
\centerline{\it Allahabad 211019, India.}

\vskip 2 true cm
\noindent {\bf Abstract}
\vskip 1 true cm

We study  coherent charge tunneling through a one-dimensional
interacting ring with a one-dimensional quantum dot embedded in one of
its arms through bosonisation. The symmetries of the effective action explain 
many of the features such as phase change between
resonances, in-phase successive
resonances and  phase-locking,  which have been observed in experiments of
coherent transport in mesoscopic rings, with a quantum dot. We also
predict changes in the  behaviour of the tunneling 
conductance in the presence of
an  Aharanov-Bohm flux through the ring. We argue that these results
hold true in general for any dot. 

\vskip 1 true cm


\newpage

Recent electron interferometry experiments\cite{Y,Z,X} on mesoscopic 
Aharanov-Bohm(AB) rings are of fundamental interest as these probe not 
only the total transmission through the resonant tunneling structure 
but also the phase associated with the electron transport. The first
such experiment by Yacoby {\it et al}\cite{Y}  on an  
AB ring with a resonant tunneling structure in the form of 
a quantum dot showed that 
there exists  a coherent component in the transport through the dot. 
Further, this
coherent transport is characterized by unusual features  - successive AB
conductance peaks are in phase and there is an abrupt change in phase
by $\pi$ when the conductance
reaches a maximum. More recent experiments 
\cite{Z,X} confirm this picture and also observe a phase 
drop of $\pi$ between successive conductance peaks. 

There have been many 
theoretical attempts\cite{ATTEMPTS} 
to explain these features. The abrupt phase change by
$\pi$ when the conductance peak reaches a maximum has been explained in terms
of the phase locking imposed by the condition that 
the two terminal conductance is an even function of the 
magnetic field\cite{YEYATI}. 
Wu {\it et al}\cite{WU} suggest that the `in phase' behaviour of 
successive conductance peaks arises due to the fact that resonant tunneling
through the whole system can be observed only when the phase shift 
introduced by the 
resonant state of the dot coincides with the transmission phase of the rest of 
the ring. Kang\cite{KANG}, in a recent work, uses the Friedel sum rule for the 
effective single particle levels in the quantum dot and a non interacting 
tight binding representation for the electrons on the ring to explain
some of these features. However, a proper understanding of the various
unusual features seen experimentally is still lacking.

Motivated by this, in this letter, we study the problem  
of coherent transport in a single channel electron ring connected to external
leads at 
$X_L$ and $X_R$ with a 1-D resonant tunneling structure
embedded in one of its arms (Fig.(1)). The single channel 
model is appropriate for a very narrow ring where one expects only a few 
1-D channels to contribute to the transmission. 
We use the bosonization approach pioneereed for transmission problems  by
Kane and Fisher\cite{KF}, and succesfully applied  
to study transport in 1-D wires in various contexts\cite{KE}.
We explain several of the distinct features characterizing coherent transport
in the interferometer device geometry in terms of
the symmetries of the theory. We also study the problem in the presence of
an external magnetic flux. This approach also allows us to study the interacting problem as well.

We begin with the tight-binding  Hamiltonian for spinless 
fermions on a ring with a hopping parameter $t$  and a short 
range repulsive Coulomb interaction $U$.
If the ring is pierced by a magnetic flux $\Phi = \int _0^L A_{\phi} dx$
where $A_{\phi}$ is the component of the 
vector potential along the ring and $L$ is the length of the
circumference of the ring, then 
the Hamiltonian can be written as 
\beq
H = t\sumjN (e ^{-{i \delta \over N}}\psi _j^{\dagger} \psi _{j+1}+h.c.)
    +U\sumjN (\psi_j^{\dagger} \psi_j)(\psi_{j+1}^{\dagger} \psi_{j+1})
\eeq
where $\delta = 2 \pi \Phi /\Phi _0$,   $\Phi _0 = hc/e$ 
is the flux quantum
and N is the total number of sites on the ring.
The fermions satisfy periodic boundary conditions on the ring - 
$\psi _{N+1} =\psi _1$.
A gauge transformation on the fermions $\psi _j 
\rightarrow e^{i \delta j \over N} \psi _j$ leads 
to the usual Hubbard form for the Hamiltonian   
\beq 
H = t \sumjN (\psi _j ^{\dagger} \psi _{j+1} + h.c) + U \sumjN ((\psi _j 
^{\dagger} \psi _j) (\psi _{j+1} ^{\dagger} \psi _{j+1}))
\eeq
with the periodic  boundary conditions on the fermions now changed to
$\psi _{N+1} = e^{i \delta } \psi _1$.
In the low-energy, long-wavelength limit, the fermion fields
can be expanded about the right and left Fermi momentum points $\pm k _F :
 \psi (x) = e^{-i k_{F} x} \psi _L (x) + e^{ i k_F x} \psi _R (x)$,
where $\psi_L(x)$ and $\psi_R(x)$ are left and right moving Fermi  fields.
Linearizing the dispersion and using the standard bosonization
technique, we 
can express, in the continuum limit,    
the fermion fields in terms of two bosonic
fields $\theta (x)$ and $\phi (x)=\int_0^x \Pi(x') dx'$ 
(where $\Pi(x)$ is the momentum of the $\theta(x)$ field) 
satisfying the commutation relations
$[\phi(x), \theta(x')] =
i \Theta(x-x')$ :
\bea
\psi _L (x) &=& e^{-i \sqrt \pi (\theta (x) -\phi (x) )}\nonumber\\
\psi _R (x) &=& e^ { i \sqrt \pi (\theta (x) + \phi (x)) }.
\label{ff}
\eea
The corresponding bosonic Hamiltonian on the ring is given by
\beq
H_{\rm ring}= v_F\int _0 ^{L} dx [{g\over 2} (\nabla\phi)^2 +
{1\over 2g} (\nabla\theta)^2]
\eeq
where $v_F$ is the Fermi velocity and $g$ is related to 
$U$ as 
$g^{-2} = ( 1+{U\over {\pi v_F}})$. 
We restrict ourselves to repulsive interactions for which $g<1$. ($g=1$
is the noninteracting limit and $g>1$ for attractive interactions.) 
In the absence of any  magnetic flux through the ring, the fermions satisfy 
periodic boundary conditions. Consistent with these
boundary conditions, one can show that the bosonic Hamiltonian is
symmetric under
$\theta(x) \longrightarrow \theta(x) + \sqpi$.
This symmetry represents the discrete particle
nature of the electrons\cite{KF}.
It can also be easily seen that a non zero magnetic flux through the
ring which couples to $\partial_{\tau} \theta(x)  = \partial_x
\phi(x)$ leaves the $\theta$ field unchanged, but
transforms the $\phi$ field as
$\phi(x) \longrightarrow \phi(x) \pm \delta/\sqpi$,
where $\delta$ is as defined earlier and $+/-$ is because 
right and left moving fermion fields  respond to the flux
in opposite ways.

\epsfxsize=4.75 in
\epsfysize=2.0 in
\begin{center}
\epsffile{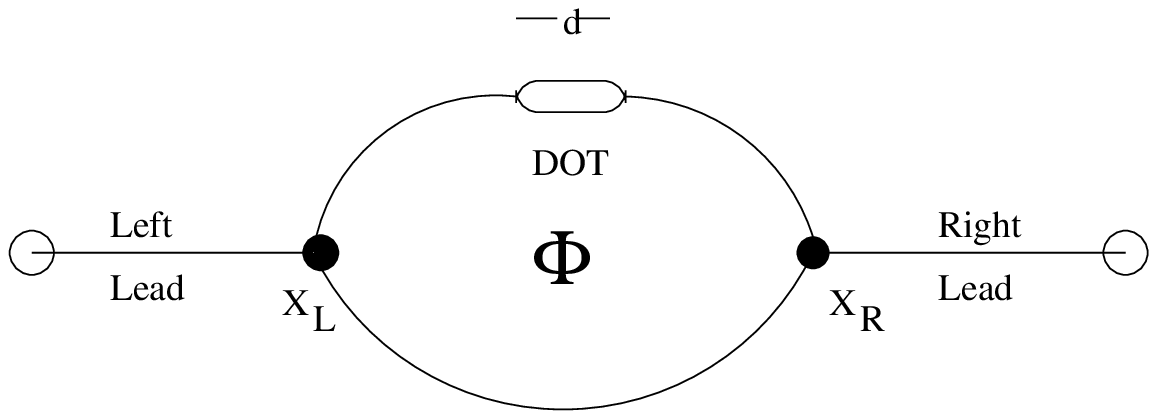}
\end{center}
\begin{itemize}
\item{ \bf Fig 1. } 
Schematic diagram of the Aharanov-Bohm ring coupled to leads on the
left and right through tunneling junctions  
and having a dot in one of its arms.
\end{itemize}

The one-dimensional  dot can be simply modelled by a symmetric 
double barrier potential,
which, in turn, can be taken to be that of two $\delta$-function  
barriers  
positioned  at $x=\frac{L}{4} -\frac{d}{2}$ and $x=\frac{L}{4}+\frac{d}{2}$
as shown in Fig. (1). 
It is appropriate to consider the large barrier limit where
Coulomb blockade effects can occur.  The effective Hamiltonian for 
the dot can then be 
written 
as\cite{KF}
\beq
H_{\rm dot}= V [\cos(2\sqpi\theta_1(\tau)-k_F d/2) + \cos(2\sqpi  
\theta_2+k_F d/2)] +V_G {\theta_2(\tau)-\theta_1(\tau)\over \sqpi}
\eeq
where $\theta_1(\tau) = \theta(x=L/4 - d/2, \tau)$, 
$\theta_2(\tau) = \theta(x= L/4 +d/2, \tau)$, $V$ is the strength
of the $\delta$-function potentials and $V_G$ is the gate voltage 
which couples
to the electrons between the two barriers. Such a system is known to 
have resonant tunneling behaviour when the Luttinger parameter 
$g>1/4$ \cite{KF}.

The leads are taken to be non-interacting and are coupled to the ring
through tunnel junctions at $X_L$ and $X_R$ as shown in Fig.(1). The lead-ring 
interaction can be described by a Hamiltonian of the form 
\beq
H_{\rm lead-ring}= [t_L b_L^{\dagger}(X_L) \psi_L(X_L) + h.c ]+
        [t_R b_R^{\dagger}(X_R) \psi_R(X_R) + h.c ]
\eeq
where $b_L$ and $b_R$ are the non-interacting fermion operators on the
left and right leads
respectively, and $\psi_{L/R}$ are the fermion operators on the ring
defined earlier.

To analyse the transport properties of the system modelled by the
action 
\beq
S= \int d \tau dx H=\int d \tau  \int_{0}^{L} dx
[H_{\rm ring} + H_{\rm dot} + H_{\rm
lead-ring}], 
\eeq
we first consider the case where there is no flux through 
the ring . Following Kane and
Fisher\cite{KF}, we see that
the quadratic degrees of freedom away from the
barriers  can be integrated out and the effective action can be
expressed as 
\beq
S_{\rm eff} ={1\over g} \sum_{i\omega_n} |\omega_n|
\{|\chi(\omega_n)|^2 +{\pi\over 4} |n(\omega_n)|^2\} + \int d \tau
{\large [} {1\over
2} {\tilde U}(n-n_0)^2 +V \cos 2\sqpi\chi\cos\pi n{\large ]} 
\eeq
with $\chi/\sqpi = (\theta_1 +\theta_2)/2\sqpi$ 
interpreted as the 
number of particles transferred across the barriers and  
$n=(\theta_2 -\theta_1)/\sqpi + k_F d/2\pi$ as the number of particles
between the two barriers. 
The first term in $V_{\rm eff}$ can be interpreted
as the energy cost to put $n$ particles in the quantum dot. 
The optimum value for the number of
particles between the barriers $n_0$ is controlled  by
$V_G$ and depends on $\tilde U=\pi \hbar v_F/g^2d$. 
The effective action is invariant  under $\chi\rightarrow
\chi + \sqpi$,
corresponding  to the transfer of an electron across the 
island or the dot with no change  
in the charge state of the dot.
However, at values of $V_G$ tuned such that the optimum value is 
$n_0=1/2$ \cite{KF}, ($V_G = {\pi \hbar v_F\over g^2d}
(-\rho_Fd+n_0)$), there is 
no extra cost to the energy to change the charge state in the dot by
one. This 
corresponds to the fact that exactly at resonance, there exists
an additional 
symmetry in the effective action,  with $\chi \rightarrow\chi + 
\sqpi/2$, along with  
a corresponding change in the charge state of the island 
$n\rightarrow n+1$  - $i.e.$ 
at each resonance, the symmetry is equivalent to changing 
  the charge state of the
island by $1$ and transferring $1/2$  
an electron
across the barrier. 
Now, let us study what happens to the fermion fields at gate 
voltages tuned such that the resonance condition is satisfied for the dot.
As we have seen above, when the dot goes through a resonance, half
an electron is transferred across the dot with a change in the charge
state of the dot by unity. Transport of half an
electron across the barrier corresponds to  
the following transformations  
for the boson fields on the upper arm of the ring -
$\theta (x) \rightarrow \theta (x) + \sqrt \pi q $ and 
$\phi(x)  \rightarrow \phi(x) \pm \sqpi q$  
with $q$  half an integer and 
$+/-$ for the $R/L$ movers in the second equation. This leads to a phase
shift of $\pi$ at each resonance. In terms of the gate voltage, the resonance
occurs at $V_G = {\pi \hbar v_F\over gd}
(-\rho_Fd+q +1/2)$. We shall call the resonances for which $q$ = even integer
as `odd' resonance and $q=$ odd integer as `even' resonance. Note that spacing 
between resonances is given by $\Delta V_G =\pi \hbar v_F/g^2d$ for the one
dimensional dot. (In general, the spacing depends on the capacitance of the dot and is given by $\Delta V_G =Q/C$, where $C$ is the capacitance of the dot. 
and is given by $\Delta V_G=Q/C$ where $C$ is the capacitance of the
dot.)

In open geometries where the Luttinger liquid  wire is connected to external 
voltage 
reservoirs, conductance experiments measure only the transmission
amplitudes which depend on the energy but not on the phase and hence,
maxima in the transmission amplitudes occur both for odd and even
resonances.  
However, in an 
interferometry geometry like that of Fig.(1), the transmission 
characteristics depend crucially on the interference patterns between the
electrons travelling through the two different paths.
For the electron travelling through the upper arm with the embedded dot, even
(odd) resonances lead to phase shifts of $2 \pi (\pi)$ respectively. 
Constructive interference with the electron travelling 
through the lower arm can therefore occur only at `even resonances',
which can 
be detected at the leads as peaks  in the conductance oscillations.
{\it So conductance maxima only occur at even resonances and 
the spacing of the gate voltages at which the maxima now occur is twice 
that observed in conductance measurements done in open geometries.} 
This implies that at the maxima, the charge state of the island or the dot
changes by even integers.   
{\it Moreover, since the phase change between two successive even
resonances is $2\pi$, 
this also explains why successive conductance maxima  are always in phase.}
Odd resonances, on the other hand,  
lead to destructive interference  with the 
electrons travelling through the lower arm and occur in between successive
conductance maxima and are characterized by a phase change of $\pi$.
{\it Thus, the phase drop of $\pi$ between successive conductance
maxima occurs because the gate voltage goes through the odd resonance
of the dot.}
One would not expect conductance maxima at these values of the
gate voltage. In fact, destructive interference at these resonances 
should make the tunneling
conductance go to zero. However, this would be true only if we 
considered only the two direct path contributions from $X_L$ to $X_R$
through the upper and lower arms in the 
path integral. But when the direct path contributions are zero, 
we must include the effects of
multipath contributions to the conductance amplitude  which lead to small but
nonzero values for the tunelling conductance.

The above analysis is valid for frequencies $\omega_n < \pi
v_F/g^2 \hbar d$, because at higher frequencies, the electron
simply sees two independent barriers and there is no resonant
tunneling. By the same reasoning, $\omega_n > \pi
v_F/g^2 \hbar (L-d)$ to ensure that the complementary distance is
sufficiently large, so that the electron sees the two barriers
sequentially and there is no resonant tunneling. 
Also, to ensure that the one-dimensional physics of the ring is
being probed, the temperatures have to be greater than $T_L =
\hbar v_F/k_B L$.

The symmetry $\chi \rightarrow
\chi + \sqpi$ is a generic symmetry of the action, whereas 
$\chi \rightarrow\chi + 
 \sqpi/2$, $n\rightarrow n+1$ is a symmetry only at
resonance. Hence, there is a phase change of $\pi$ at each successive 
resonance,
where the charge state of the island changes by unity. However, 
in between resonances, the symmetry is
restored to $\chi \rightarrow
\chi + \sqpi$ which does not allow phase changes. {\it Phase rigidity
is thus a consequence of this symmetry.}
What about the scale over which the symmetry changes or the width of
the resonances? The naive expectation for the width of the resonance
is that it be of the same order of the energy scale in the problem,
which is the dot energy scale or $\hbar v_F/g^2 d$. However, for
mesoscopic systems, the system size offers another energy scale of the
order of $1/L$, which is much smaller than $1/d$ in the limit where
this analysis is valid and can lead to much narrower
resonances. Moreover, for interacting systems in one dimension, it is
well-known that resonances are extremely narrow, degenerating into
$\delta$- function peaks as $T\rightarrow 0$. Hence, we suggest that
{\it the extra-ordinary abruptness of the phase change on
scales much smaller than $k_B T$ can
be explained by a combination of two facts. One is that the
the resonance is related to the symmetry
of the fermions on the ring, whose relevant energy scale is given by $1/L$.
The second is that for interacting electrons, resonances
are extremely narrow. In fact, a study of the scale over which the
phase change takes place is equivalent to the study of the resonance
line shape for the resonance peak. Thus, we predict that the width 
over which $\pi$
changes should be the same as the width of 
the resonance maxima.
The measurement of this width should thus be a measurement of the
Luttinger parameter $g$ in the one-dimensional wire}\cite{us}.

When we introduce flux through the ring, we see that
the symmetries on the bosonic fields at even and odd
resonances through the upper arm are also changed to
$\theta (x) \rightarrow \theta (x) + \sqrt \pi q $ and   
$\phi (x) \rightarrow \phi(x) \pm \sqpi q \pm \delta/\sqpi$,
where $q$ is integer or half-integer for even and odd resonances
and $+/-$ is for the $R/L$ movers respectively. For the
lower arm, they are given by
$\theta (x) \rightarrow \theta (x) + \sqrt \pi q $ and   
$\phi (x) \rightarrow \phi(x) \pm \sqpi q \pm \delta/\sqpi$.
But here, $q$ is always an integer. 

Now let us consider some particular cases.
\begin{itemize}

\item{} $\delta = \pi$

At odd resonances, both 
$\psi _{L}(x)$ and $\psi_R(x)$ acquire a phase shift 
$\pi$ 
as they  travel through the dot. But they also acquire a phase shift
of $\pi$ due to the flux.  Hence, in this case, 
there is a destructive interference 
with the corresponding fields from the lower arm when $q$ is an
integer - $i.e$ at even resonances.  
When $q$ is odd,  
the fermions through the upper and lower arms interfere
constructively to give rise to conductance peaks. Thus, the position
of the conductance maxima shift to the position of the minima when
there was no flux.

\item{} $\delta = \pi/2$

When one-quarter fluxes are introduced, there is yet another
twist which comes into play. Precisely at $\delta=\pi/2$, it becomes
possible to have constructive intereference between left-moving
electrons through the upper arm with right moving holes through the
lower arm and vice-versa. The phases of the left and right movers at
even resonance after one full circuit through the ring are given by
\bea
\psi_L \rightarrow e^{-i\pi/2}\psi_L,&\quad& \psi_R\rightarrow
e^{i\pi/2}\psi_R \quad {\rm with  ~~dot} \nonumber \\
\psi_L^C \rightarrow e^{i\pi/2}\psi_L^C,&\quad& \psi_R^C\rightarrow
e^{-i\pi/2}\psi_R^C \quad {\rm without ~~dot}
\eea
and at odd resonance by
\bea
\psi_L \rightarrow e^{i\pi/2}\psi_L,&\quad& \psi_R\rightarrow
e^{-i\pi/2}\psi_R \quad {\rm with ~~ dot} \nonumber \\
\psi_L^C \rightarrow e^{3i\pi/2}\psi_L^C,&\quad& \psi_R^C\rightarrow
e^{-3i\pi/2}\psi_R^C \quad {\rm without ~~dot}
\eea
In both cases, it is easy to see that $\psi_L$ through the upper arm
and $\psi_R^C$ through the lower arm have the
same phases and interfere constructively and so do 
$\psi_R$ and $\psi_L^C$. Also, this constructive interference happens
both for the odd and even resonances and there should be maxima
in the transmission conductance for both cases. 
Hence,  the spacing between conductance peaks should be halved as
compared to the spacing without any flux. 
This  halving of the spacing   of the conductance 
maxima at these values of $\delta$ has also been noted by
Kang\cite{KANG},  who
computed the tunneling conductance explicitly using the Friedel sum
rule for phase change through the dot. However, here we
understand the reason why the odd resonances survive at these
particular values of the external flux in terms of the 
symmetries of the theory.

\item{}$\delta$ = arbitrary

Here, as for the case when $\delta=\pi$, we still expect the
transmission at even resonances, where there is no phase shift through
the dot, to interfere constructively and lead to conductance
maxima. However, the position of the conductance maxima shift
continuously as a function of the flux.

\end{itemize}

Note that the entire  analysis has no dependence  on the value of the 
Luttinger parameter 
$g$  except that it be within the range where resonant tunneling behaviour is
allowed. The only difference that one expects 
between $g=1$ and $g\ne 1$ is in the
widths of the regions of the phase change, and the 
widths of the transmission maxima.  At $T=0$,
and in the thermodynamic limit, for $g\ne 1$, the resonance peaks are
expected to be infinitely sharp. However, for finite $T$ and for
mesoscopic lengths $L$, one expects the widths to have appropriate power
law dependences on both these quantities. Whether the abrupt nature of
the phase change is related to the well-known fact that
interactions appear to narrow resonances, or whether it depends only
on the fact that the scale over which the resonance occurs is related
to the ring energy scales rather than the dot energy scales, is a more
detailed question, which needs the explicit computation of the
tunneling conductance and the line shapes\cite{us}.

In conclusion, in the above analysis, 
we have explicitly used the effective action for
a 1-D resonant tunneling structure to show that the effect of going through
the dot resonant state leads to a change in phase for the electron fields
which can  then be evaluated.   
But a similar analysis 
is also valid  for a general dot, as long as the dot is embedded in 
a narrow wire where the single channel approximation holds good.
The dot can be thought of as  providing  effective single particle 
energy levels for resonance, 
($Q^2/C$, where $C$ is the capacitance of the dot), 
as well as a phase change whenever an odd
number of fermions tunnel through it. Transmission through the dot can
then be
thought of in terms of hopping between a local impurity situated at the site 
of the dot and the electron fields \cite{WONG}. The effect of such a local 
impurity on the fermions in the ring 
is to cause phase shifts which in the bosonic 
representation can be expressed  in terms of transformations  
on the associated boson fields. A more detailed analysis in terms of
boundary conformal field theory and an explicit computation of the tunneling
conductance will be reported elsewhere\cite{us}. 
  
Thus, we have been able to explain many of the distinct features seen in
experiments on coherent transport through a mesoscopic ring with a dot
embedded in one of its arms, through the symmetries of the effective
action for the coherent transport. 
Conductance maxima occur only at even resonances which allow
for constructive interference between the two different paths and 
since the same symmetry exists at all even resonances, this  also 
explains why successive maxima  are always in phase. Between two succesive
maxima, there
is an odd resonance which is characterized by a phase drop of $\pi$.
However, note that this means that the two transmission maxima are separated
from one another by the addition of two electrons to the dot, whereas
experiments seem to indicate that it is more likely that the maxima are
separated by the addition of a single electron. To put it another way,
in general, one would not expect the Coulomb blockade minima to
coincide with the minima expected due to destructive interference.

The phase rigidity between
phase changes is explained as a consequence of the symmetry of the action which
corresponds to the discrete particle nature of the electrons.
The abruptness of the phase change can be  related both 
to the fact that the underlying
symmetry
change occurs over electron energy scales and not over dot energy
scales and also to the narrowness of the
resonances expected for interacting fermions. Within this picture, in
fact, it is harder to understand why the width of the conductance
maxima follows the standard non-interacting Breit-Wigner form. In
other words, the conductance maxima behave as if single electrons are
tunneling through, whereas the minima behave as if the system is
interacting. This is still a puzzle that has to be understood better. 
 
In the presence of an Aharanov-Bohm flux through the 
ring, the positions of the resonances are shifted.
We also see a period
doubling in the case when the flux through the ring is one-quarter of
the flux quantum, in agreement with the result of Kang.
At a more detailed level,
the non-zero tunneling conductance amplitude at the 'odd resonances'
require multi-path contributions or contributions of higher
dimensional operators in the path integral. 
We expect our results  to hold even for a general dot. At resonance,
the effect of the dot is expected to be that of 
a local impurity on the fermions in the ring, leading to phase
shifts.  

\section*{Acknowledgments}
P.D is grateful to C.S.I.R. (India) for financial support and also thanks
M.R.I., Allahabad for hospitality during the course of this work.

\end{document}